\documentclass[11pt]{article}
\usepackage{graphicx}

\input epsf

\newcommand{\be}{\begin{equation}}
\newcommand{\ee}{\end{equation}}
\newcommand{\bea}{\begin{array}}
\newcommand{\ea}{\end{array}}
\newcommand{\beqa}{\begin{eqnarray}}
\newcommand{\eeqa}{\end{eqnarray}}
\newcommand{\bean}{\begin{eqnarray*}}
\newcommand{\eean}{\end{eqnarray*}}

\def\up#1{\leavevmode \raise.16ex\hbox{#1}}

\def\sqr#1#2{{\vcenter{\vbox{\hrule height.#2pt
        \hbox{\vrule width.#2pt height#1pt \kern#1pt
          \vrule width.#2pt}
        \hrule height.#2pt}}}}

\newcommand{\gapproxeq}{\lower .7ex\hbox{$\;\stackrel{\textstyle
>}{\sim}\;$}}
\newcommand{\lapproxeq}{\lower .7ex\hbox{$\;\stackrel{\textstyle
<}{\sim}\;$}}

\newcounter{appendice}

\def\thebibliography#1{{\bf REFERENCES\markboth
 {REFERENCES}{REFERENCES}}\list
 {[\arabic{enumi}]}{\settowidth\labelwidth{[#1]}\leftmargin\labelwidth
 \advance\leftmargin\labelsep
 \usecounter{enumi}}
 \def\newblock{\hskip .11em plus .33em minus -.07em}
 \sloppy
 \sfcode`\.=1000\relax}

\begin{document}

\centerline{ \LARGE Remarks on an Exact Seiberg-Witten map }

\vskip 2cm

\centerline{ {\sc    A. Stern}  }

\vskip 1cm
\begin{center}

{\it Department of Physics, University of Alabama,\\
Tuscaloosa, Alabama 35487, USA}

\end{center}

\vskip 2cm

\vspace*{5mm}

\normalsize
\centerline{\bf ABSTRACT} 

\vspace*{5mm}
We obtain the leading derivative corrections to an expression for
 the Seiberg-Witten map  given by Banerjee and Yang and show how they  affect the noncommutative deformation of the Maxwell action, as well as the matter coupling in noncommutative emergent gravity. 
\newpage

\scrollmode 

The Seiberg-Witten map relates noncommutative gauge potentials, fields and transformation parameters  to their commutative counterparts.\cite{Seiberg:1999vs}
It is defined such that  gauge transformations in the commutative theory induce  gauge transformations  in the corresponding noncommutative theory.  This is the  Seiberg-Witten consistency condition.  Exact solutions \cite{Okuyama:1999ig},\cite{Liu:2000mja},\cite{Mukhi:2001vx},\cite{Jurco:2001my},\cite{Okawa:2001mv},\cite{Liu:2001pk}, as well as series expansions  in terms of the noncommutative parameters
$\theta^{\mu\nu}$ (see for example \cite{Ulker:2007fm}) have been obtained for the map.  They have  often been applied for the purpose of finding
corrections to standard electromagnetic theory and to general relativity (for example \cite{Guralnik:2001ax},\cite{Berrino:2002ss},\cite{Stern:2007an},\cite{Fabi:2008ta}).  However, as
the results  are generally rather technically involved, the map has rarely been applied beyond the second order in $\theta^{\mu\nu}$.  Since $\theta^{\mu\nu}$ has units of length-squared, the corrections are expansions in $\theta^{\mu\nu}/$length-squared, and so results obtained so far are generally  only valid for length scales much larger than $\sqrt{\theta^{\mu\nu}}$. 

There exists, on the other hand,  an extremely  simple expression for the map, known in \cite{Liu:2000mja}, which was discussed by Banerjee and Yang\cite{Banerjee:2004rs},\cite{Yang:2004vd} and referred to as exact.  It relates commutative   field strengths in one coordinate system to their noncommutative counterparts in another coordinate system - without involving derivatives of the fields.   The map was applied in \cite{Banerjee:2004rs}, \cite{Yang:2004vd} to give a nonlinear deformation of electrodynamics and an  emergent  metric for gravity.  Maxwell equations result in the former, along with  $local$ relations between the usual four vector fields $\vec D$, $\vec H$, $\vec E$ and $\vec B$, and  describe an effective noncommutative medium. In the context of noncommutative  emergent gravity\cite{Rivelles:2002ez}, \cite{Steinacker:2007dq},\cite{Grosse:2008xr},\cite{Yang:2008fb},  
 Banerjee and Yang recovered the standard minimal  coupling of matter fields to gravity. More specifically, for the case of a scalar field $\phi$ in four dimensional space-time, all interactions with the metric tensor $g_{\mu\nu}$ are contained in the action
\be -\frac 12\int d^4x \sqrt{-g} g^{\mu\nu} \partial_\mu\phi \partial_\nu\phi \label{grvtsclrcplng}\ee   However, it is well known \cite{Seiberg:1999vs},\cite{Liu:2000mja},
\cite{Mukhi:2001vx},\cite{Jurco:2001my},\cite{Yang:2006dk} that the map for the fields involves  derivative corrections, and so strictly speaking, the results in \cite{Banerjee:2004rs}, \cite{Yang:2004vd} can only be considered exact for  constant field strengths.  Just as they have been known to contribute to D-brane actions\cite{Okawa:1999cm},\cite{Wyllard:2000qe},\cite{Das:2001xy}, the higher derivative corrections will affect  the noncommutative deformation of electrodynamics and noncommutative emergent gravity.  So for the former, the exact  Lagrangian cannot be written as a local function of the fields, while for the latter, the exact matter coupling should contain infinitely many higher derivative corrections to (\ref{grvtsclrcplng}). 
In this note we obtain the leading order derivative corrections for these two dynamical systems.  The procedure to be used is simply to demand that the Seiberg-Witten consistency condition is satisfied up to this order.  It can be extended  to arbitrary order, although this procedure most likely will not lead to  simple recursion relations.

We begin by showing that the map  discussed in  \cite{Banerjee:2004rs},\cite{Yang:2004vd} (which is equivalent to equations (\ref{xitozta}), (\ref{FeqlG}) and (\ref{ninxctsw}) in this article) fails to satisfy the Seiberg-Witten consistency condition starting at third order in $\theta^{\mu\nu}$.
We specialize to $U(1)$ gauge theory and utilize 
 the
Groenewold-Moyal star product\cite{groe},\cite{moy}
\be \star = \exp\;\biggl\{ \frac {i}2 \theta^{\mu\nu}\overleftarrow{
  \frac{\partial}{\partial \xi^\mu }}\;\overrightarrow{ \frac{\partial}{\partial \xi^\nu }} \biggr\} \;,\label{gmstr} \ee 
 where here $\theta^{\mu\nu}=-\theta^{\nu\mu}$ are constants and
  $\overleftarrow{\frac{\partial}{\partial \xi^\mu }}$ and $\overrightarrow{ \frac{\partial}{\partial \xi^\mu }}$ are left and right
derivatives,
respectively, with respect to some coordinates $\xi^\mu,\; \mu=0,1,2,3$, which parametrize the space-time manifold.
The noncommutative gauge theory is described by potentials $\hat A_{\mu}$ and field strengths   $\hat F_{\mu\nu} $ 
\be \hat F_{\mu\nu}(\xi)= \frac{\partial}{\partial \xi_\mu }\hat A_\nu (\xi)- \frac{\partial}{\partial \xi_\nu } \hat A_\mu(\xi)
-i[\hat A_\mu(\xi), \hat A_\nu(\xi)]_\star \;,\label{fldstrng}\ee   
 where $[\;,\;]_\star$ denotes the star commutator, i.e.,  
$[f,g]_\star\equiv f\star g - g\star f$ for any two function $f$ and $g$.  So for example,  $[\xi_\mu ,\xi_\nu]_\star= i\theta^{\mu\nu}$.  The 
Groenewold-Moyal star implies that only odd powers of $\theta^{\mu\nu}$ appear in the expansion of the star commutator
\be [f,g]_\star\; =\; i\theta^{\mu\nu}\frac{\partial f}{\partial \xi^\mu }\frac{\partial g}{\partial \xi^\nu }\;-\;\frac i{24}\theta^{\mu\nu}\theta^{\rho\sigma}\theta^{\kappa\lambda}\frac{\partial^3 f}{\partial \xi^\mu \partial \xi^\rho\partial \xi^\kappa}\frac{\partial^3 g}{\partial \xi^\nu \partial \xi^\sigma\partial \xi^\lambda}\;+\;{\cal O}(\theta^5)
\ee
The noncommutative potentials and fields have gauge variations 
\beqa \delta \hat A_\mu(\xi) &=&  \frac{\partial}{\partial \xi^\mu }\hat\Lambda(\xi) + i[ \hat\Lambda(\xi),\hat A_\mu(\xi) ]_\star\label{ncgvrtsA}\\ & &\cr
 \delta \hat F_{\mu\nu}(\xi) &=& i[ \hat\Lambda(\xi),\hat F_{\mu\nu}(\xi) ]_\star\;, \label{ncgvrtsF}\eeqa
 where $\hat \Lambda(\xi)$ are infinitesimal parameters.

The Seiberg-Witten map relates $\hat A_\mu,\;\hat F_{\mu\nu}$ and $ \hat\Lambda$ to their commutative counterparts,  $ A_\mu,\; F_{\mu\nu}=\partial_\mu A_\nu -\partial_\nu A_\mu$ and $ \Lambda$, respectively.
The expression   in \cite{Banerjee:2004rs},\cite{Yang:2004vd} gives the noncommutative fields $\hat F_{\mu\nu}$ evaluated for coordinates $\xi^\mu$ in terms of the commutative  fields $F_{\mu\nu}$ evaluated for coordinates $x^\mu$,
where  $x^\mu$ and $\xi^\mu$ are related via the noncommutative potential
\be x^\mu(\xi) =\xi^\mu + \theta^{\mu\nu}\hat  A_\nu(\xi)  \label{xitozta}\ee 
Thus
\be \hat F_{\mu\nu}(\xi) = {\cal G}_{\mu\nu}(x(\xi))\;,\label{FeqlG}\ee 
and so 
\be \frac {\partial \hat F_{\mu\nu}}{\partial \xi^\mu } = \frac {\partial {\cal G}_{\mu\nu}}{\partial x^\sigma} \frac {\partial x^\sigma}{\partial \xi^\mu }\label{rlfprtldrvs} \ee
According to \cite{Banerjee:2004rs},\cite{Yang:2004vd}, ${\cal G}_{\mu\nu}$ depends only on $F_{\mu\nu}$  (and $\theta^{\mu\nu}$), and not its derivatives.  Specifically, 
\be  {\cal G}_{\mu\nu}(x)=F_{\mu\rho}(x)\biggl[\frac 1{1 +\theta F(x)}\biggr]^\rho_{\;\;\;\nu} \;\;\;\label{ninxctsw}\ee 
This was shown to be valid for constant fields $F_{\mu\nu}$ to all orders in $\theta$.\cite{Liu:2000mja}  For nonconstant  $F_{\mu\nu}$ one can  easily compare the commutative and noncommutative fields  evaluated in a common coordinate system, upon  expanding in $\theta$
\beqa \hat F_{\mu\nu} &=& F_{\mu\nu} + F_{\mu\nu}^{(1)} + F_{\mu\nu}^{(2)} +\cdots \cr & &\cr 
{\cal G}_{\mu\nu} &=& F_{\mu\nu} -(F\theta F)_{\mu\nu}+(F\theta F\theta F)_{\mu\nu}+\cdots \cr & &\cr x^\mu& =&\xi^\mu + \theta^{\mu\nu}\Bigl( A_\nu +  A_\nu^{(1)} +\cdots \Bigr)\;,\eeqa
where  $A_{\mu}^{(n)}$ and  $F_{\mu\nu}^{(n)}$ denote, respectively,  terms in the noncommutative potentials and field strengths of $n^{\rm th}$ order in $\theta$.
So for example
\beqa F_{\mu\nu}^{(1)}&=&-(F\theta F)_{\mu\nu} + \partial_\rho F_{\mu\nu} (\theta A)^{\rho}   \cr & &\cr
 F_{\mu\nu}^{(2)}&=&(F\theta F\theta F)_{\mu\nu}-\partial_\rho(F\theta F)_{\mu\nu}(\theta A)^\rho + \partial_\rho F_{\mu\nu} (\theta A^{(1)})^\rho +\frac 12 \partial_\rho\partial_\sigma F_{\mu\nu}(\theta A)^\rho (\theta A)^\sigma \label{ordrxpndsn}\;
\eeqa
 These results agree with known solutions for the Seiberg-Witten map up to second order in $\theta$.  (See for example \cite{Stern:2008wi}.) 
Below we show that (\ref{ninxctsw}) cannot be relied upon for nonconstant $F_{\mu\nu}$ beyond the second order, or more generally, that any ${\cal G}_{\mu\nu}$  expressed in terms of only $F_{\mu\nu}$  does not satisfy the Seiberg-Witten consistency conditions beyond the second order.

Since  ${\cal G}_{\mu\nu}$ is a function only of the commutative field strengths it is   gauge invariant.  On the other hand, from (\ref{xitozta}) and (\ref{ncgvrtsA}), the coordinates $x^\mu$ of the domain of ${\cal G}_{\mu\nu}$  have  nonvanishing gauge variations,
\be \delta x^\mu(\xi) =i [ \hat\Lambda(\xi),x^{\mu}(\xi) ]_\star \ee
Infinitesimal gauge variations of the right hand side of (\ref{FeqlG}) are then
\beqa \delta {\cal G}_{\mu\nu}(x)&=& i\frac {\partial {\cal G}_{\mu\nu}}{\partial x^\eta} \; [ \hat\Lambda,x^{\eta} ]_\star \cr & &\label{xpndrhs}\\
&=& \frac {\partial {\cal G}_{\mu\nu}}{\partial x^\eta} \;\biggl(-\theta^{\alpha\beta}\frac{\partial \hat \Lambda}{\partial \xi^\alpha }\frac{\partial x^\eta}{\partial \xi^\beta }\;+\;\frac 1{24}\theta^{\alpha\beta}\theta^{\rho\sigma}\theta^{\kappa\lambda}\frac{\partial^3 \hat \Lambda}{\partial \xi^\alpha \partial \xi^\rho\partial \xi^\kappa}\frac{\partial^3 x^\eta}{\partial \xi^\beta \partial \xi^\sigma\partial \xi^\lambda}\;+\;{\cal O}(\theta^5) \biggr)
\nonumber
\eeqa
We wish to compare this to infinitesimal gauge variations of the left hand side of (\ref{FeqlG})  given by (\ref{ncgvrtsF}), 
\beqa 
 \delta \hat F_{\mu\nu}(\xi) &=& -\theta^{\alpha\beta}\frac{\partial \hat\Lambda}{\partial \xi^\alpha }\frac{\partial \hat F_{\mu\nu}}{\partial \xi^\beta }\;+\;\frac 1{24}\theta^{\alpha\beta}\theta^{\rho\sigma}\theta^{\kappa\lambda}\frac{\partial^3 \hat\Lambda}{\partial \xi^\alpha \partial \xi^\rho\partial \xi^\kappa}\frac{\partial^3 \hat F_{\mu\nu}}{\partial \xi^\beta \partial \xi^\sigma\partial \xi^\lambda}\;+\;{\cal O}(\theta^5)\label{xpndlhs}\eeqa
Using (\ref{rlfprtldrvs}),  the   terms proportional to $ \frac{\partial \hat\Lambda}{\partial \xi^\alpha }$ in (\ref{xpndrhs})
 and (\ref{xpndlhs}) agree, but the terms proportional to $\frac{\partial^3 \hat\Lambda}{\partial \xi^\alpha \partial \xi^\rho\partial \xi^\kappa}$ do not.
In (\ref{xpndlhs}) the latter contribute at third order in $\theta$, while they contribute at fourth order in (\ref{xpndrhs}) [since $\frac{\partial^3 x^\eta}{\partial \xi^\beta \partial \xi^\sigma\partial \xi^\lambda}$ goes like $\theta$ to leading order].  Therefore (\ref{FeqlG}) can only be trusted up to second order in $\theta$, and moreover, the proof did not require knowledge of the explicit expression for $\mathcal G_{\mu\nu}$ in terms of $F_{\mu\nu}$ given in (\ref{ninxctsw}).

 The exact result for ${\cal G}_{\mu\nu}$ in (\ref{FeqlG}) must contain infinitely many derivatives of the field strengths.  We can easily obtain the leading order derivative corrections to (\ref{ninxctsw}) and to the noncommutative deformation of the Maxwell action. 
 In order for (\ref{xpndrhs}) and (\ref{xpndlhs}) to agree at third order in $\theta$, we have to add a term  to (\ref{ninxctsw})  which at leading order  has the gauge variation
 \be 
 \frac 1{24}\theta^{\alpha\beta}\theta^{\rho\sigma}\theta^{\kappa\lambda}\;\partial_\alpha \partial _\rho\partial_\kappa \Lambda \; \partial_\beta \partial_\sigma\partial_\lambda F_{\mu\nu}\ee
 So if we now replace (\ref{ninxctsw}) by  
 \beqa {\cal G}_{\mu\nu} &=& F_{\mu\nu} -(F\theta F)_{\mu\nu}+(F\theta F\theta F)_{\mu\nu}\cr & &\cr & &-(F\theta F\theta F\theta F)_{\mu\nu}+\frac 1{24}\theta^{\alpha\beta}\theta^{\rho\sigma}\theta^{\kappa\lambda}\partial_\alpha\partial_\rho A_\kappa\partial_\beta\partial_\sigma\partial_\lambda  F_{\mu\nu}+{\cal O}(\theta^4)\label{Gup23rdr}\eeqa
 we  get a corrected solution to the Seiberg-Witten equations which is valid up to third order in $\theta$ and which now involves derivatives of the field strength.
 From (\ref{FeqlG}), the action for  noncommutative $U(1)$ gauge theory in four dimensional space-time
\be  {\cal S}_F = - \frac 14 \int d^4\xi\;\hat F_{\mu\nu}(\xi)\star \hat F^{\mu\nu}(\xi)\label{ncactn}\ee can be re-expressed as an action for  commutative fields
  \be {\cal S}_F =  - \frac 14 \int d^4 x \Big|\frac {\partial \xi}{\partial x}\Big|  \;{\cal G}_{\mu\nu}(x){\cal G}^{\mu\nu}(x) \;,\label{ncglgrn}\ee since for the Groenewold-Moyal star $\int d^4\xi\;\alpha(\xi)\star\beta(\xi)=\int d^4\xi\;\alpha(\xi)\beta(\xi)$, for well behaved functions  $\alpha(\xi)$ and $\beta(\xi)$.  Then (\ref{Gup23rdr}) leads to the presence of terms in the action which involve derivatives of the commutative  field strength. The leading third order contribution of such terms can be expressed as
\be \frac 1{192}\int d^4 x\; \theta^{\alpha\beta}\theta^{\rho\sigma}\theta^{\kappa\lambda}\;\partial_\alpha\partial_\rho F_{\kappa\lambda}\;\partial_\beta
F_{\mu\nu}\;\partial_\sigma F^{\mu\nu}\label{3rdrdrem} \ee
For this we did not need to consider the Jacobian factor $|\frac {\partial \xi}{\partial x}|$.
Higher derivative terms from the Jacobian factor can also contribute, but these occur at higher order in $\theta$.  This is more easily seen in two  dimensions where the inverse Jacobian is simply \be\Big|\frac {\partial x}{\partial \xi}\Big|=1+\theta^{01}\biggl[ \frac{\partial}{\partial \xi_0 }\hat A_1 (\xi)- \frac{\partial}{\partial \xi_1 } \hat A_0(\xi)
-i[\hat A_0(\xi), \hat A_1(\xi)]_1  \biggr]\;,\ee where $[\;,\; ]_1 $ denotes the first order term of the star commutator.  Up to third order in $\theta$, one can approximate the right hand side by  $1+\theta^{01}\hat F_{01}$, and upon substituting in (\ref{Gup23rdr}), one gets that the leading higher derivative term contributes at fourth order in $\theta$.

 In \cite{Rivelles:2002ez}  the Seiberg-Witten map was used to transform  noncommutative $U(1)$ gauge theory into  an effective theory of gravity, or noncommutative emergent gravity.  For this one  can couple the gauge theory to a scalar field and obtain the Seiberg-Witten map from the commutative to the noncommutative covariant derivative of the field, which we denote,  respectively, by $\partial_\mu\phi$ and $\hat D_\mu\hat \phi$.    Here the noncommutative scalar field $\hat \phi$ is assumed to be in the adjoint representation of the gauge group, with  \be \hat D_\mu\hat \phi (\xi)=  \frac{\partial}{\partial \xi^\mu }\hat\phi(\xi) + i[ \hat\phi(\xi),\hat A_\mu(\xi) ]_\star \;,\ee while the commutative field $\phi$ is gauge invariant. 
 A simple expression  for the Seiberg-Witten map was given in  \cite{Banerjee:2004rs}, which is of the form 
  \be \hat D_\mu\hat \phi (\xi)= {\cal D}_{\mu}\phi(x(\xi))\;,\label{DeqlG}\ee 
  where ${\cal D}_{\mu}\phi$ is assumed to  depend only on $\partial \phi$ and $F$.  Specifically, 
\beqa  {\cal D}_{\mu}\phi &=&\biggl[\frac 1{1 +F\theta}\biggr]_\mu^{\;\;\;\nu}\partial_\nu\phi\cr & &\cr&=& (1-F\theta+ F\theta F\theta+\cdots)_\mu^{\;\;\;\nu}\partial_\nu\phi \;\;\;\label{Dinxctsw}\eeqa 
This was obtained using a compactification of (\ref{ninxctsw}).
 We can use it to compare $\hat D_\mu\hat \phi$ and $\partial \phi$ evaluated in a common coordinate system. Expanding  up to second order in $\theta$,  
 \be \hat D_\mu \hat \phi = \partial_\mu \phi  + [D_\mu \phi]^{(1)} + [D_\mu \phi]^{(2)}+\cdots\;\;\;\label{frrdrfcd}\ee
 Then
 \beqa 
 [D_\mu \phi]^{(1)} &=&-(F\theta)_\mu^{\;\;\nu}\partial_\nu\phi +(\theta A)^\nu\partial_\mu\partial_\nu\phi 
\cr & &\cr
[D_\mu \phi]^{(2)} &=& (F\theta F\theta)_\mu^{\;\;\nu}\partial_\nu\phi
-\partial_\rho (F\theta)_\mu^{\;\;\nu} (\theta A)^\rho \partial_\nu\phi- (F\theta)_\mu^{\;\;\nu} (\theta A)^\rho \partial_\rho\partial_\nu\phi \cr & &\cr & & + (\theta A^{(1)})^\rho \partial_\mu\partial_\rho \phi +\;\frac 12(\theta A)^\rho (\theta A)^\sigma\partial_\mu \partial_\rho \partial_\sigma \phi 
\label{obocvrntdrv}\;,\eeqa
which can be verified to satisfy the Seiberg-Witten consistency condition up to this order.
Unfortunately,    (\ref{Dinxctsw}) cannot be trusted beyond second order in $\theta$.  Following the previous arguments, we get  a mismatch of the gauge variations of the left hand side of   (\ref{DeqlG}), i.e.,
 \be  \delta\hat D_\mu\hat \phi =i [ \hat\Lambda, \hat D_\mu\hat \phi ]_\star\;, \ee
 with those of the right hand side
\be \delta {\cal D}_{\mu}\phi= i\frac {\partial {\cal D}_{\mu}\phi}{\partial x^\eta} \; [ \hat\Lambda,x^{\eta} ]_\star \ee 

As before, the exact result for the Seiberg-Witten map must contain infinitely many derivatives of the fields.
 In order to obtain the correct result up to third order in $\theta$, we need to add a term  to ${\cal D}_{\mu}\phi$  in (\ref{Dinxctsw}) which  has the gauge variation
 \be 
 \frac 1{24}\theta^{\alpha\beta}\theta^{\rho\sigma}\theta^{\kappa\lambda}\;\partial_\alpha \partial _\rho\partial_\kappa \Lambda \; \partial_\beta \partial_\sigma\partial_\lambda \partial _{\mu}\phi\;,\ee and so up to  third order,  ${\cal D}_\mu\phi$ can take the form
 \beqa  {\cal D}_{\mu}\phi &=& (1-F\theta+ F\theta F\theta- F\theta F\theta F\theta + \cdots)_\mu^{\;\;\;\nu}\partial_\nu\phi\cr & &\cr & & + \frac 1{24}\theta^{\alpha\beta}\theta^{\rho\sigma}\theta^{\kappa\lambda}\;\partial_\alpha \partial _\rho A_\kappa \; \partial_\beta \partial_\sigma\partial_\lambda \partial _{\mu}\phi + \cdots\;\;\;\label{3rdrdrsftm}\eeqa Using the map, the action for  noncommutative scalar field in four dimensional space-time
\be  {\cal S}_\phi = - \frac 12 \int d^4\xi\;\hat D_{\mu}\hat \phi(\xi)\star \hat D^{\mu}\hat \phi(\xi)\ee can be re-expressed as an action for the commutative fields $\phi$ \be {\cal S}_\phi  =  - \frac 12 \int d^4 x \Big|\frac {\partial \xi}{\partial x}\Big|  \;{\cal D}_{\mu}\phi(x){\cal D}^{\mu}\phi(x) \;\label{sfncglgrn}\ee  Then (\ref{3rdrdrsftm}) implies the presence of terms in the action which involve higher derivatives of the  scalar field, which to leading  order can be written
\be \frac 1{192}\int d^4 x\; \theta^{\alpha\beta}\theta^{\rho\sigma}\theta^{\kappa\lambda}\;\partial_\alpha\partial_\rho F_{\kappa\lambda} \;\partial_\beta
\partial_{\mu}\phi \;\partial_\sigma \partial^{\mu}\phi\;,\ee
where again we did not need to consider the Jacobian factor at this order.
Clearly,  terms such as these cannot be absorbed in the standard coupling of the scalar field to gravity (\ref{grvtsclrcplng}).  Thus, as before, higher derivative terms appear beginning  at third order in $\theta$.  They represent  additional and novel contributions to the dynamics for noncommutative emergent gravity theories.\cite{Rivelles:2002ez},\cite{Steinacker:2007dq},\cite{Grosse:2008xr},\cite{Yang:2008fb}

\bigskip
\noindent
{\bf Acknowledgment}

\noindent
I am grateful to H. Steinacker  for useful discussions and to H. S. Yang for pointing out reference \cite{Yang:2006dk}.

 \end{document}